\documentclass[aps,prl,twocolumn]{revtex4}
\usepackage{bm,color,amsmath,amssymb,mathrsfs,latexsym,graphicx,psfrag}

\newcommand{\up}{\uparrow}
\newcommand{\dw}{\downarrow}

\newcommand{\bk}{{\mathbf k}}
\newcommand{\bq}{{\mathbf q}}

\newcommand{\be}{\begin{equation}}
\newcommand{\ee}{\end{equation}}

\def\bea{\begin{eqnarray}}
\def\eea{\end{eqnarray}}

\begin{document}

\title{Topological semimetals and topological insulators in rare earth monopnictides}
\author{Minggang Zeng$^{\dag1,2}$, Chen Fang$^{\dag3}$, Guoqing Chang$^{1,2}$, Yu-An Chen$^{3}$, Timothy Hsieh$^3$, Arun Bansil$^4$, Hsin Lin$^{\ast1,2}$ and Liang Fu$^{\ast\ast3}$}
\affiliation{$^1$Centre for Advanced 2D Materials and Graphene Research Centre, National University of Singapore, Singapore 117546}
\affiliation{$^2$Department of Physics, National University of Singapore, Singapore 117542}
\affiliation{$^3$Department of Physics, Massachusetts Institute of Technology, Cambridge, Massachusetts 02139, USA}
\affiliation{$^4$Department of Physics, Northeastern University, Boston, Massachusetts 02115, USA}
\begin{abstract}
{We use first principles calculations to study the electronic properties of rock salt rare earth monopnictides La$X$ ($X=$N, P, As, Sb, Bi). A new type of topological band crossing termed `linked nodal rings' is found in LaN when the small spin-orbital coupling (SOC) on nitrogen orbitals is neglected. Turning on SOC gaps the nodal rings at all but two points, which remain gapless due to $C_4$-symmetry and leads to a 3D Dirac semimetal. Interestingly, unlike LaN, compounds with other elements in the pnictogen group are found to be topological insulators (TIs), as a result of band reordering due to the increased lattice constant as well as the enhanced SOC on the pnictogen atom. These TI compounds exhibit multi-valley surface Dirac cones at three $\bar{M}$-points on the $(111)$-surface.}
\end{abstract}
\maketitle

The discovery of three-dimensional (3D) Dirac semimetals and Weyl semimetals, both theoretically\cite{YoungPRL2012,WangPRB2012,WangPRB2013,WanPRB2011,BurkovPRL2011,GoPRL2012,Lu2013,Weng2015,Huang2015} and experimentally\cite{Liu2014,LiuS2014,NeupaneNc2014Observation,Xu2015,xu2013arxiv,Lu2015,Xu2015a,Lv2015}, attracts great interest and effort into the emergent field of topological semimetals (TSM). A $d$-dimensional topological semimetals have Fermi surfaces of reduced dimensions below $d-1$.
In 3D, the Fermi surface of a topological semimetal is constituted of  points or lines instead of surfaces. The new types of Fermi surfaces is a result of robust band crossings, or topological band crossings, between the conduction and the valence bands, protected by topology and/or symmetry. Topological semimetals have distinct physical properties such as surface Fermi arcs, negative magnetoresistence and topologically nontrivial spin texture near the Fermi surface.

With the only exception of 3D Weyl semimetals, the stability of all TSM phases requires the presence of symmetries in addition to lattice translation, such as point group or time-reversal\cite{YoungPRL2012, WangPRB2012,WangPRB2013,YangPrl2013Theory}. Breaking these symmetries in different ways typically drives the system into distinct topological phases: breaking TRS in a 2D Dirac semimetal, e.g., graphene, leads to a quantum anomalous Hall state; breaking spin rotation in the same system leads to a 2D TI; breaking time-reversal or 3D inversion a 3D Dirac semimetal splits a Dirac point into a pair of Weyl nodes resulting in a Weyl semimetal\cite{MurakamiNJoP2007,Choapa2011Possible, DasPRB2013,DoraPRB2013,HalaszPRB2012,HouPRL2013}; breaking rotation symmetry in a 3D Dirac semimetal results in a 3D TI.
Therefore, a TSM could be viewed as the parent state of many interesting phases\cite{Liu2014}.

\begin{figure}[tbp]
\includegraphics[width=0.45\textwidth]{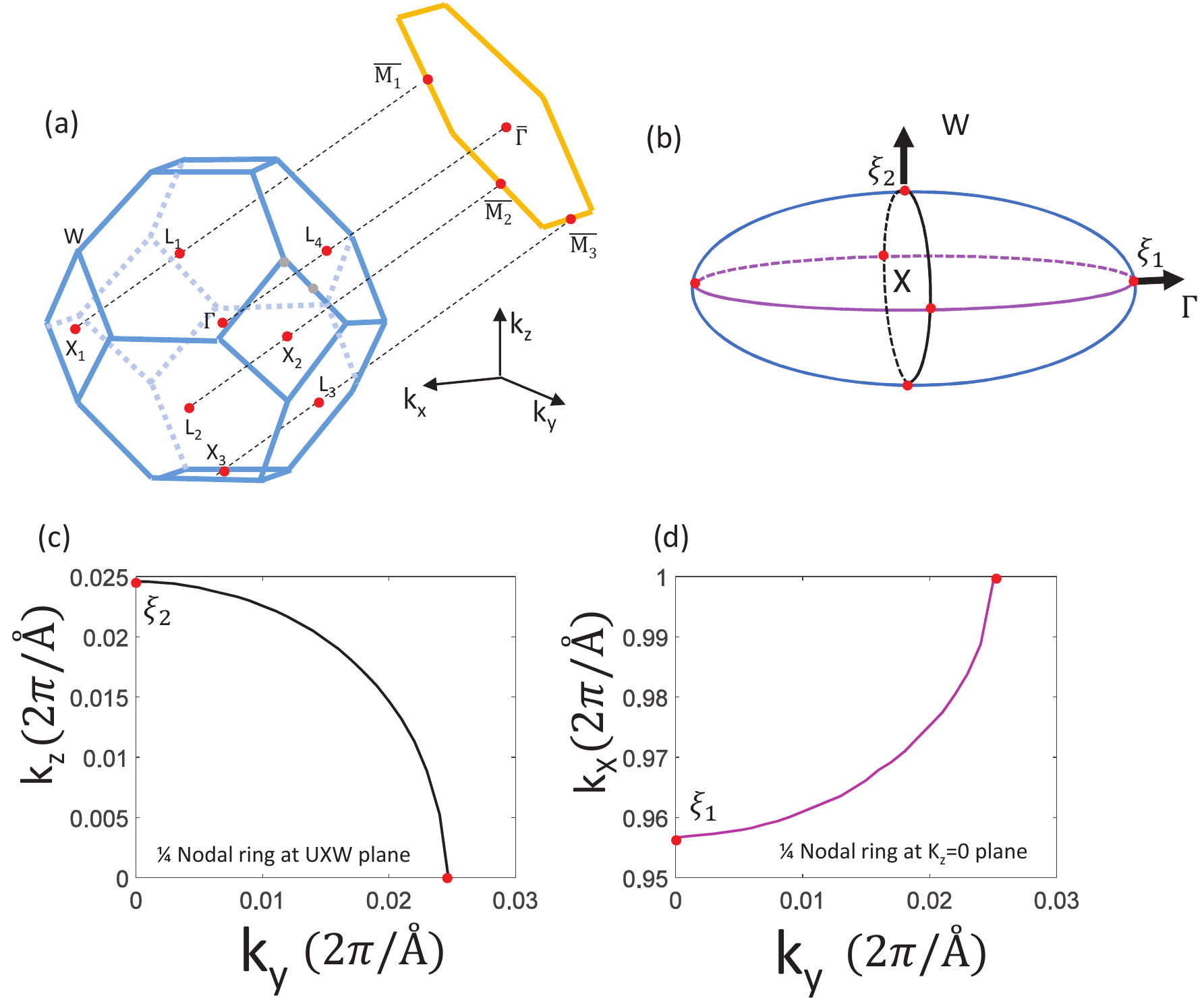}
\caption{(a) The 3D Brillouin zone of the face-centered cubic lattice and its projection to the surface Brillouin zone of the $(111)$-surface. (b) A schematic of the three nodal rings around one $X$-point in LaN. (c,d) The calculated configuration of the nodal rings on the $yz$-plane and the $xy$-plane, respectively. Due to symmetry, only one quarter of each ring is shown.}
\label{fig:1}
\end{figure}

In this work, based on first-principles calculations, we predict topological semimetal and topological insulator phases in a family of lanthanum monopnictides with a simple rock salt structure: La$X$, where $X=$N, P, As, Sb and Bi. When the small SOC is neglected in LaN, the system is an exotic TSM: the band crossing points form three intersecting nodal rings, which look like the equator and two perpendicular longitudes of a football centered at each X point, shown in Fig.\ref{fig:1}(b) (see also Ref.\onlinecite{Weng2014} for a similar configuration of nodal lines proposed in 3D graphene networks). These nodal rings are protected by three mirror planes as well as spin rotation symmetry, and as the latter is broken by a perturbative SOC, they are gapped almost everywhere, leaving two Dirac points behind. Hence in LaN, born from the nodal ring semimetal is a Dirac semimetal with six Dirac points, two near each $X$-point, in the Brillouin zone (BZ). 3D Dirac semimetals have been experimentally identified in Na$_3$Bi and Cd$_3$As$_2$\cite{LiuS2014,NeupaneNc2014Observation,Xu2015,xu2013arxiv}. Considering that the former is chemically unstable under ambient conditions, and the latter has a very complicated crystal structure, we note that LaN, being a simple binary compound with rock salt structure, has certain advantages from materials perspective. In the other compounds ($X$=P, As, Sb and Bi), the bulk is a 3D TI with full direct gap. A band inversion between lanthanum $d$-orbitals and pnictogen $p$-orbitals at $X$-point appears in all five compounds, and hence is \emph{not} the reason for the topological phase transition from TSM to TI. Our study shows that a significant increase in the lattice constant changes the orbital nature of the valence band from $p_x$-orbital to a linear combination of $p_{y,z}$-orbitals; this change in orbital nature causes the topological distinction between a TSM in LaN and a TI in the other monopnictides. To demonstrate the TI phase, we calculate the band structure of a thick slab of LaBi normal to the $[111]$-direction, finding three surface Dirac cones near three $\bar{M}$s in the surface Brillouin zone (SBZ).

Our first-principles calculations are implemented in the VASP package with the generalized gradient approximation (GGA) and the projector augmented wave (PAW) method \cite{PAYNE1992,ROBERTSON1991,PERDEW1992}. Lattice constants of rock salt lanthanum monopictnides are adopted from experimental values \cite{OlceseJoPFMP1979,OnoJotLM1972,YOSHIHARAJOTLM1975,PALEWSKIPSSAR1993,LEGERJOPCSP1984}. A Monkhorst-Pack k-mesh (11 $\times$ 11 $\times$ 11) is used to sample the Brillouin zone \cite{MONKHORST1976}. Fig.\ref{fig:1}(a) shows the BZ of the lanthanum monopnictides. To account for the electron-electron interaction, a mean field Hubbard correction term ($U$) is introduced into the frame of density functional theory\cite{Dudarev1998} (DFT+U). The absence of $f$-electrons and the significant dispersion of $5d$-states in all compounds suggest the itinerant nature of the conduction band. In accordance with this observations, it is found in our calculation that when $U<0.5$eV, the lattice constants obtained are consistent with the experimental values. We hence choose $U=0.25$eV for all calculations performed in this work. We also note that there is no qualitative difference between the results for $U=0$ and $U=0.5$eV.

\begin{figure}[tbp]
\includegraphics[width=0.45\textwidth]{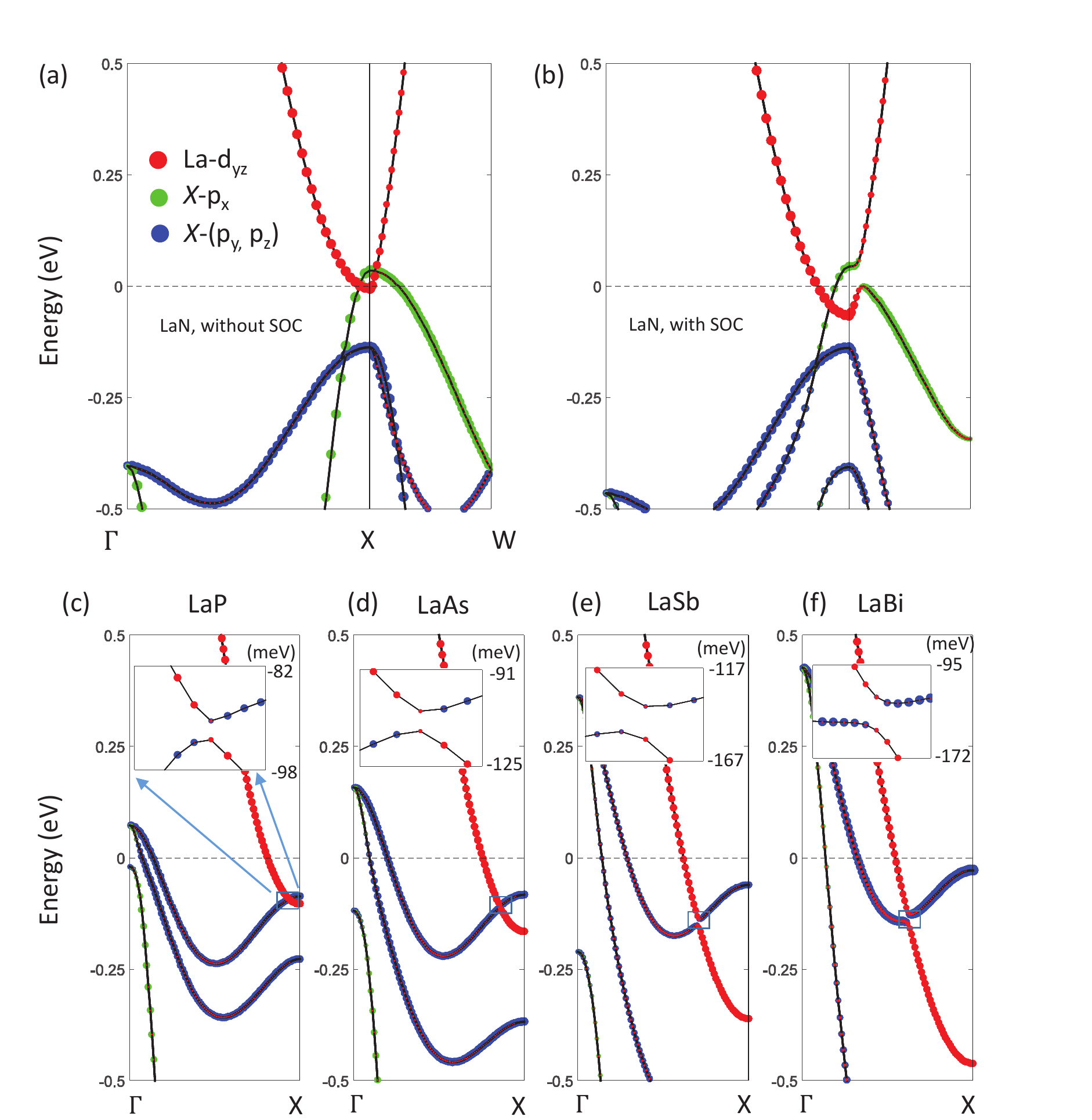}
\caption{Band structures of La$X$ with orbital analysis. (a,b)The band structure of LaN in the absence and the presence of SOC, respectively. (c-f) The band structures of LaP, LaAs, LaSb and LaBi along $\Gamma{X}$, respectively, in the presence of SOC, with insets showing details of the anti-crossing.}
\label{fig:2}
\end{figure}

The band structure along high-symmetry lines for LaN in the absence of SOC is plotted in Fig.\ref{fig:2}(a). A band inversion at $X$ is seen. The $p_x$ states of N is about 45meV higher than the $d_{yz}$ states of La. Due to the opposite parity of the two orbitals, the band inversion would have indicated a 3D TI if the direct gap were nonzero. However, there are two band crossing points near X: one along $\Gamma{X}$ and the other along $XW$. A symmetry analysis shows that neither of them is a discrete crossing point in BZ, but each is an intersection of two nodal rings, protected by mirror symmetries. There are three mirror planes: $M_{yz}$ that maps $x$ to $-x$, $M_{xz}$ that maps $y$ to $-y$ and $M_{xy}$ that maps $z$ to $-z$. Let us focus on $M_{xy}$ for now. In the BZ, bands on the plane defined by $k_z=0$ can be labeled by the eigenvalues of $M_{xy}$, $m_{xy}=\pm1$. A band with $m_{xy}=+1$ cannot anti-cross another band with $m_{xy}=-1$, because any hybridization would break mirror symmetry. This is exactly our case: $p_x$-orbital is invariant under $m_{xy}$ thus having $m_{xy}=+1$, while $d_{yz}$-orbital has $m_{xy}=-1$. Hence, on the $k_z=0$-plane, the conduction band and valence band cross each other in a nodal ring, passing through $\xi_1$. Similar analysis proceeds for the $M_{xz}$ mirror plane and we derive another nodal ring on the $k_y=0$-plane also passing through $\xi_1$. Therefore, $\xi_1$ is the intersection of two nodal rings. We can similarly deduce that $\xi_2$ is also such an intersection of two nodal rings, protected by $M_{xz}$ and $M_{yz}$ respectively. From symmetry analysis, we have already deduced three nodal rings that are all centered at $X$ and cross each other at $\xi_{1,2}$ and their symmetry equivalents, as shown in the schematic in Fig.\ref{fig:1}(b). An extensive DFT calculation away from high-symmetry lines confirms this prediction, and the calculated configurations of the nodal rings are found in Fig.\ref{fig:1}(c,d).

We substantiate the above analysis by deriving a minimal $k\cdot{p}$-Hamiltonian around $X$-point. Without SOC, there are two bands at $X$-point, so the $k\cdot{p}$ is given by a two-by-two matrix as function of $\bq\equiv\bk-X$, which can be decomposed as the sum of the identity matrix and three Pauli matrices with $\bq$ dependent coefficients:
\bea\label{eq:h}
h(\bq)=\sum_{\mu=0,x,y,z}d_\mu(\bq)\sigma_\mu.
\eea
Since we are interested in the band crossing, the $\sigma_0$-term, which represents an overall kinetic energy, can be neglected henceforth. The little group of $X$-point, formed by all symmetry operations that leave $X$-point invariant, gives symmetry constraints on the functional forms of $d_{i=x,y,z}(\bq)$. The little group is $D_{4h}\otimes\{I,T\}$, where $\{I,T\}$ is the group generated by time-reversal operation, $T$. The generators of the little group include: fourfold rotation about $x$-axis $C_{4x}$, mirror plane $M_{xy}$, 3D inversion $P$ and time-reversal $T$. The symmetry constraints hence take the form
\bea\label{eq:general_constraints}
C_{4x}h(q_x,q_y,q_z)C^\dag_{4x}&=&h(q_x,-q_z,q_y),\\
\nonumber
M_{xy}h(q_x,q_y,q_z)M_{xy}^\dag&=&h(q_x,q_y,-q_z),\\
\nonumber
Ph(\bq)P^\dag&=&h(-\bq),\\
\nonumber
Th(\bq)T^\dag&=&h^\ast(-\bq),
\eea
where in the last equation we have used the fact that time-reversal is anti-unitary. If the basis vectors are chosen to be $|p_x\rangle$ for isospin up and $|d_{yz}\rangle$ for isospin down, the symmetry operations correspond to the following matrices: $C_{4x}=\sigma_z$, $M_{xy}=\sigma_z$, $P=-\sigma_z$ and $T=\sigma_0$. Substituting these into Eqs.(\ref{eq:general_constraints}), we obtain the following constraints for $d_i(\bq)$'s:
\bea\label{eq:d}
d_x&=&0,\\
\nonumber
d_y(\bq)&=&uq_xq_yq_z+O(q^5),\\
\nonumber
d_z(\bq)&=&m-uq_x^2-v(q_y^2+q_z^2)+O(q^4).
\eea
The dispersion of Eq.(\ref{eq:h}) is $\sqrt{d_x^2+d_y^2+d_z^2}$, so the nodal points are determined by $d_x=d_y=d_z=0$. The explicit forms of $d_i(\bq)$'s given in Eqs.(\ref{eq:d}) immediately yield three nodal rings: one circle given by $q_x=0$, $q_y^2+q_z^2=m/v$, one ellipse given by $q_y=0$, $uq_x^2+vq_z^2=m$ and another ellipse given by $q_z=0$, $uq_x^2+vq_y^2=m$.

The linked nodal rings are, however, generically unstable against perturbative SOC, i.e., the conduction and the valence bands anti-cross, despite the unbroken mirror symmetry. This is because the above mentioned mirror symmetries, which only act in the spatial degrees of freedom, are no longer symmetries of the system, as the spatial degrees of freedom are now coupled to the spin degrees of freedom by SOC. A `real' mirror symmetry acts simultaneously on the spatial and the spin spaces: $M_{xy}$, for example, not only sends $z$ to $-z$, but also sends $s_{x,y}$ to $-s_{x,y}$, i.e., performs a $\pi$-rotation about $y$-axis in the spin space, because spin is a pseudo vector. With the additional spin-rotation, each band with mirror eigenvalue $m_\pm$ in the non-SOC system becomes two degenerate bands with mirror eigenvalues $+im_{\pm}$ and $-im_\pm$ for the spin-up and spin-down sub-bands, respectively. Therefore, the band crossing between $p_x$ band and $d_{yz}$ band on $k_z=0$-plane is no longer protected by $M_{xy}$: the spin-up (spin-down) sub-band of $p_x$ has the \emph{same} eigenvalue of $M_{xy}$ as the spin-down (spin-up) sub-band of $d_{yz}$ and they will anti-cross. The nodal rings hence disappear generically. However, the crossing point $\xi_1$ along $\Gamma{X}$ remains gapless for another symmetry reason. The high-symmetry line $\Gamma{X}$ is a $C_{4x}$-invariant line, meaning that each band along this line can be labeled by its $C_{4x}$ eigenvalue. With SOC, $C_{4x}$ is composed of fourfold rotations in both the spatial and the spin spaces. According to this definition, the two sub-bands of the $p_x$ band have $C_{4x}$ eigenvalues of $e^{\pm{i}\pi/4}$, and those of the $d_{yz}$ band have eigenvalues of $-e^{\pm{i}\pi/4}$. Therefore, the two doublet bands can still cross each other as they have different $C_4$-eigenvalues and the four-band crossing point, $\xi_1$, as well as its time-reversal equivalent, are two 3D Dirac points. In Fig.\ref{fig:2}(b), it is confirmed in calculation that while the crossing at $\xi_2$ opens a gap, $\xi_1$ remains gapless.

The analysis can also be made explicit in the $k\cdot{p}$-theory. $h(\bq)$ in Eq.(\ref{eq:h}) only involves the orbital degrees of freedom, and we need to find the terms coupling the orbital and the spin degrees of freedom, i.e., SOC terms, allowed by the little groups. As explained above, all point group operations (except for 3D inversion which does not act on spin) are now constituted of a spatial part and a spin rotation. One should also remember that time-reversal acts nontrivially in the spin space, sending $\{|\up\rangle,|\dw\rangle\}$ to $\{|\dw\rangle,-|\up\rangle\}$. If the basis vectors are chosen to be $\{|p_x,\up\rangle,|p_x,\dw\rangle,|d_{yz},\up\rangle,|d_{yz},\dw\rangle\}$, the little group generators take the form: $C_{4x}=\sigma_z\exp(-is_x\pi/4)$, $M_{xy}=-i\sigma_zs_z$, $P=\sigma_z$ and $T=K(is_y)$, where $s_{x,y,z}$ are Pauli matrices acting on the real spin, in contrast to $\sigma_{x,y,z}$ that act on the isospin. Under the symmetry constraints in Eqs.(\ref{eq:general_constraints}), we find the following SOC terms are allowed
\bea
H_{soc}=\lambda\sigma_x(q_ys_y-q_zs_z)+O(q^3).
\eea
With SOC terms added, the dispersion becomes
\bea
E=\sqrt{[m-uq_x^2-v(q_y^2+q_z^2)]^2+\lambda^2(q_y^2+q_z^2)},
\eea
whose only band crossing points are given by $q_y=q_z=0$ and $q_x^2=m/u$.

Now we move down the pnictogen in the table of element, from N to P and further down to As, Sb and Bi. We notice that the lattice constant increases drastically from 5.3{\AA} to 6.03{\AA} in LaP, then gradually increases up to 6.58{\AA} in LaBi. This change modifies the bands near $X$-point significantly: it pushes the pnictogen $p_x$-band down in energy by a large amount, such that the valence band becomes the $p_{y,z}$ doublet (without SOC). Upon adding SOC, the $p_{y,z}$-bands (now four bands including spin) split into a lower band of $J_z=\pm1/2$, consisting of $|p_y+ip_z,\dw\rangle$ and $|p_y-ip_z,\up\rangle$ state and a higher band of $J_z=\pm3/2$ consisting of $|p_y+ip_z,\up\rangle$ and $|p_y-ip_z,\dw\rangle$ state. As a combined result of lattice and SOC, we find that while the band inversion also happens at $X$-point between the La $d$-states and the pnictogen $p$-states, the band crossing along $\Gamma{X}$ becomes an anti-crossing. See Fig.\ref{fig:2}(c-f). The gap at the anti-crossing point is as small as $\sim$3meV in LaP, increasing to 6meV, 20meV and 35meV in LaAs, LaSb and LaBi, respectively. The anti-crossing can again be understood in a symmetry analysis. Per the definition of above, the $J_z=\pm3/2$ states have $C_4$-eigenvalues of $-e^{\pm{i}\pi/4}$, same as those of the $d_{yz}$-states: $C_4$-symmetry can no longer protect their crossing. Here we see that the change in the orbital nature of the valence band induces a topological phase transition from a Dirac semimetal to a 3D TI.

An effective $k\cdot{p}$ theory can be established to confirm this result and be used to gain more information. If we the four basis vectors are $\{|p_y+ip_z,\rightarrow\rangle, |p_y-ip_z,\leftarrow\rangle, |d_{yz},\rightarrow\rangle, |d_{yz},\leftarrow\rangle$, where $\rightarrow/\leftarrow$ means the spin is along positive/negative $x$-directions, the generators of the little group are represented by $C_{4x}=\exp(-i3\sigma_zs_x\pi/4)$, $M_{xy}=i\sigma_zs_z$, $P=-\sigma_z$ and $T=-K(i\sigma_0s_y)$. The generic form (up to the first order of $\bq$) of the $k\cdot{p}$ theory is
\bea\label{eq:5}
H(\bq)=M\sigma_z+r_1q_x\sigma_xs_x+r_2\sigma_y(s_yq_y-s_zq_z).
\eea
The dispersion of this Hamiltonian is fully gapped, and since the parity of the occupied states at $\bq=0$ changes as $M$ changes sign, we see that a band inversion at $X$ changes the strong $Z_2$ index of the system according to the Fu-Kane formula\cite{FuPRB2007}. The explicit form in Eq.(\ref{eq:5}) also enables us to calculate the change in the mirror Chern number\cite{Teo2008,Hsieh2012} when the band inversion happens, i.e., when $M$ changes sign. There are three independent mirror planes to consider: $M_{xy}$, $M_{yz}$ and $M_{011}$, where $M_{011}$ is the plane rotated from $M_{yz}$ by 45 degrees about $x$-axis. Note that although $M_{xy}$ and $M_{yz}$ are related by $C_{4z}$ in the whole BZ, they are \emph{not} related by any little group operation at $X$-point, and are hence independent operations. A straightforward calculation shows that the change in the three mirror Chern numbers, as $m$ changes from positive to negative, are given by
\bea
\Delta{C}_{xy}&=&\Delta{C}_{011}=1,
\\\nonumber
\Delta{C}_{yz}&=&sign(r_1r_2).
\eea

While La$X$ is predicted to be a 3D TI when $X=$P, As, Sb and Bi, a single surface Dirac cone may only exist on certain surface terminations, due to the large hole pocket near $\Gamma$. To see this point, let us consider the $(001)$-surface. Since there are an odd number of surface Dirac cones, there must be one Dirac cone at $\bar{M}$ or at $\bar{\Gamma}$ in the SBZ. At the same time, we notice that in the 3D BZ, the line projecting to $\bar{M}$ passes two $X$-points, thus having two band inversions. Therefore a single Dirac cone cannot exist at $\bar{M}$. $\bar\Gamma$-point in SBZ is the projection of the line $\Gamma{X}$ in BZ, having only one band inversion, but in Fig.\ref{fig:2}(c-f), we see that along this line the indirect gap is closed, i.e., collapsing the bands to one point results in a continuous spectrum. Therefore the single Dirac cone is buried inside the bulk projection continuum and hence cannot be observed.

\begin{figure}[tbp]
\includegraphics[width=0.45\textwidth]{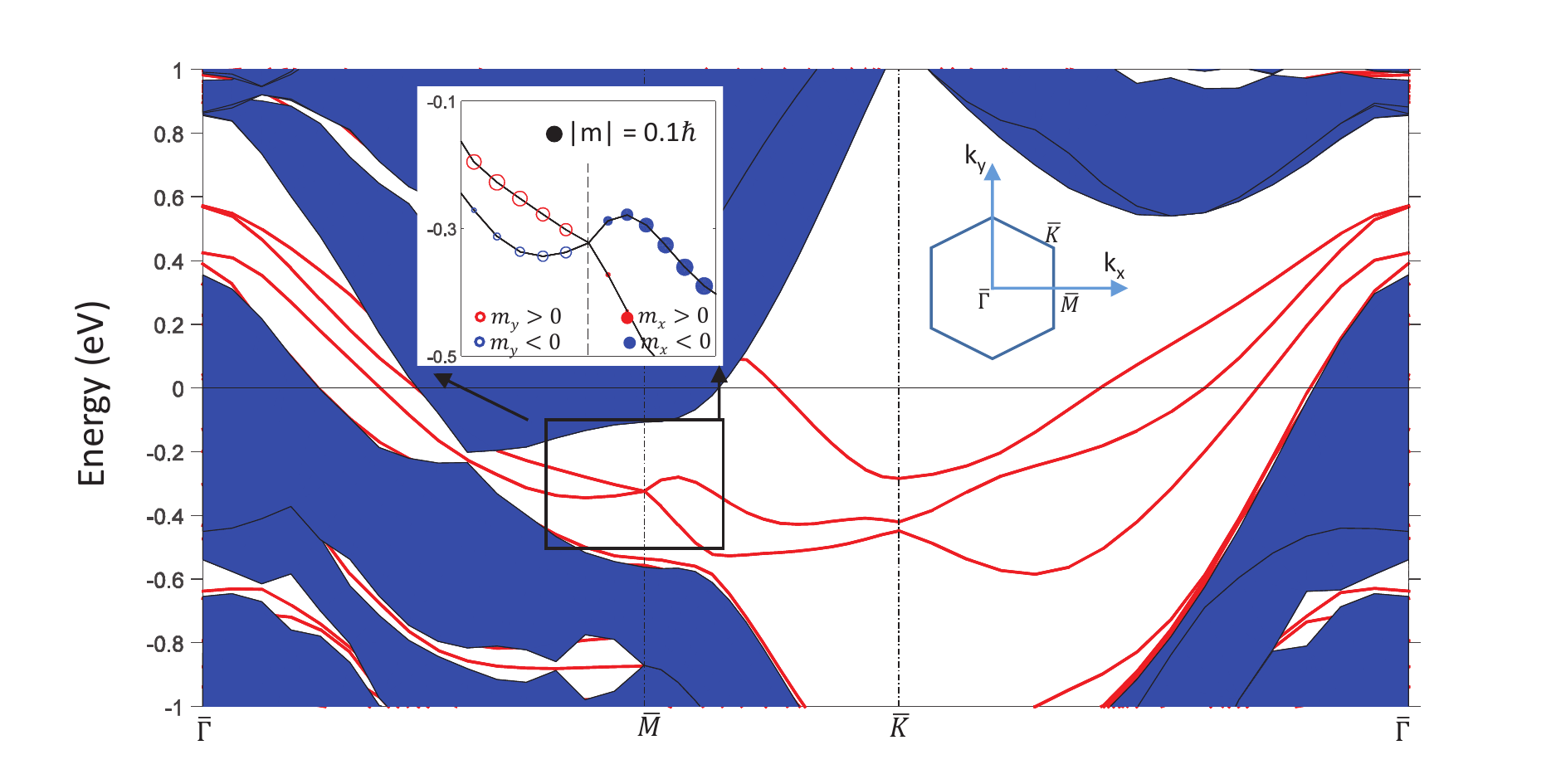}
\caption{The band structure on the $(111)$-surface of a 20-layer slab of LaBi. The inset shows the spin texture near $\bar{M}$.}
\label{fig:3}
\end{figure}

So far we have argued that no single Dirac cone appears on the $(001)$-surface. The topological surface state can, however, be observed on the $(111)$-surface. The four time-reversal invariant momenta are $\bar\Gamma$ and three $\bar{M}$'s, and either there is one Dirac cone at $\bar\Gamma$ or there are three Dirac cones at three $\bar{M}$'s. In Fig.\ref{fig:1}(a), we notice the following projections from lines in BZ to points in SBZ: $\Gamma{L}_4\rightarrow\bar\Gamma$ and ${X}_i{L}_i\rightarrow\bar{M}_i$. The band inversion happens at all $X$, so according to the projection, we expect three Dirac cones at three $\bar{M}$'s. The prediction is supported by calculation on a 20-layer LaBi slab normal to the $[111]$-direction. In Fig.\ref{fig:3}, we see a spin-split Dirac cone at $\bar{M}$. Due to the spin-orbital coupling, the spin degrees of freedom are entangled with orbital degrees of freedom, and therefore while the surface bands are spin-split, the physical spins are not fully polarized.

To conclude, we theoretically predict topological semimetals as well as topological insulators in lanthanum monopnictides with rock salt structure, La$X$ ($X=$N, P, As, Sb and Bi). We identify LaN as a nodal ring semimetal when spin-orbital coupling is neglected; and starting from this phase, we use analysis and numerics to show that it becomes a 3D Dirac point when perturbative spin-orbital coupling is taken into account. Moving down in the pnictogen column, we find LaP, LaAs, LaSb and LaBi to be 3D topological insulators. Our calculation shows that the topological transition from Dirac semimetal to topological insulator is induced by the change in the orbital character of the valence band caused by an increase in the lattice constant. We further argue that the topological surface states of these 3D topological insulators cannot be observed on the certain surfaces such as $(001)$-surface but may be observed on the $(111)$-surface. We calculate the surface states in a LaBi $(111)$-slab as demonstration.

\acknowledgements{The work at National University of Singapore is supported by the National Research Foundation, Prime Minister's Office,  Singapore under its NRF fellowship (NRF Award No. NRF-NRFF2013-03). The work at Massachusetts Institute of Technology is supported by the Science and Technology Center for Integrated Quantum Materials, NSF Grant No. DMR-1231319 (CF), and the DOE Office of Basic Energy Sciences, Division of Materials Sciences and Engineering under Award No. DE-SC0010526 (LF).}

$^\dag$These two authors contribute equally to this work.\\
$^\ast$nilnish@gmail.com\\
$^{\ast\ast}$liangfu@mit.edu

\end{document}